\documentclass[twocolumn,a4paper,superscriptaddress,floatfix,longbibliography]{revtex4}
\usepackage[english]{babel}
\usepackage[utf8]{inputenc}
\usepackage[normalem]{ulem}
\usepackage{amsmath}
\usepackage{graphicx,epstopdf}
\usepackage[colorinlistoftodos]{todonotes}
\usepackage{amssymb,epsfig,color,textcase}
\usepackage{hyperref}
\usepackage{doi}
\usepackage[title]{appendix}

\def\blue{\color{blue}}

\begin{document}

 \title{Interplay of the Jahn--Teller effect and spin-orbit coupling: The case of trigonal vibrations}

\author{Sergey V.~Streltsov$^*$}
\affiliation{Institute of Metal Physics, S. Kovalevskoy St. 18, 620990 Ekaterinburg, Russia}
\affiliation{Department of Theoretical Physics and Applied Mathematics, Ural Federal University, Mira St. 19, 620002 Ekaterinburg, Russia}
\email{streltsov@imp.uran.ru}

\author{Fedor V. Temnikov}
\affiliation{Institute of Metal Physics, S. Kovalevskoy St. 18, 620990 Ekaterinburg, Russia}

\author{Kliment I.~Kugel}
\affiliation{Institute for Theoretical and Applied Electrodynamics, Russian Academy of Sciences, Izhorskaya str. 13, 125412 Moscow, Russia}
\affiliation{National Research University Higher School of Economics, 101000 Moscow, Russia}

\author{Daniel I. Khomskii}
\affiliation{II. Physikalisches Institut, Universit$\ddot a$t zu K$\ddot o$ln,
Z$\ddot u$lpicher Stra$\ss$e 77, D-50937 K$\ddot o$ln, Germany}

\date{\today}

\begin{abstract}
We study an interplay between the orbital degeneracy and the spin-orbit coupling (SOC) giving rise to spin-orbital entangled states  in concentrated systems (cooperative Jahn-Teller (JT) effect). As a specific example, we analyze the interaction of electrons occupying triply degenerate single-ion $t_{2g}$ levels with trigonal vibrations (the $t\otimes T$ problem). A more general problem of the electron--lattice interaction involving both tetragonal and trigonal vibrations is also considered. It is shown that the result of such interaction crucially depends on the occupation of $t_{2g}$ levels leading to either the suppression or the enhancement of the JT effect by the SOC.

DOI:  \href{https://doi.org/10.1103/PhysRevB.105.205142}{\blue 10.1103/PhysRevB.105.205142}
\end{abstract}

\maketitle

\section{Introduction}

The effects related to spin-orbit coupling (SOC) have recently become quite topical especially due to their decisive role in the physics of topological  insulators and other topological materials. These effects are also important in such strongly correlated electron systems as 4$d$ and 5$d$ transition metal compounds. In contrast with 3$d$ compounds, the large SOC characteristic of 4$d$ and 5$d$ transition metal ions can play a dominant role in the formation of electron structure determining the sequence and multiplet characteristics of the energy levels. Therefore, in such systems, we are dealing with the spin-orbit entangled electron states  \cite{Takayama2021}. This means that the spin and orbital degrees of freedom become intermixed leading to a more pronounced contribution of magnetism to the orbital characteristics.

Indeed, the orbital degeneracy, leading in particular to the Jahn--Teller (JT) effect, is quite common in many transition metal compounds. Until recently, it was predominantly studied in 3$d$ systems containing such well-known JT ions as Mn$^{3+}$ and Cu$^{2+}$. Currently, however,  the  attention is gradually shifting to the study of 4$d$ and 5$d$ compounds. In this case, the SOC starts to play a more and more important role. Therefore a question arises: what is the concerted outcome of the JT effect and strong SOC? The most natural expectation is that SOC would suppress the JT effect. Indeed, due to JT distortions, the orbital degeneracy is lifted, and it becomes favorable to put an electron at the state with a real wave function, with a particular quadrupole moment. At the same time, the SOC rather prefers the states with complex wave functions. Note here that even the first, rather old treatment \cite{Opik1957a} has revealed that in the simplest case of one electron per site the JT effect is gradually suppressed with increasing SOC (characterized by the SOC constant $\lambda$).

In 3$d$ systems we usually deal with the high-spin state (satisfying the first  Hund's rule stabilizing the state with maximum possible spin), in which case we often have a situation with partially filled $e_g$ states (like those in Mn$^{3+}$ or Cu$^{2+}$). However, for $e_g$ electrons, the SOC is in the first approximation quenched. In contrast, for 4$d$ and 5$d$ systems, we typically have  low-spin states with very often partial filling of triply-degenerate $t_{2g}$ orbitals, but for these, the SOC is not quenched, and just in this case, the most realistic for 4$d$ and 5$d$ systems, one should expect an important role of SOC.

At the same time, in many cases, there still remains an orbital degeneracy even if the SOC is very strong. The orbital degeneracy typically manifests itself in the involvement of the crystal lattice occurring in the form of vibronic interactions, i.e. those related to the JT effect   \cite{Bersuker1989,Bersuker2006,khomskii2014transition,Streltsov-UFN,Streltsov2020-chem}.

Such a strong interplay of electronic and lattice characteristics in the systems with spin-orbit entangled states should lead to a plethora of novel quantum phenomena, the analysis of which now seems to be only at the initial stage. In this connection, let us note some early  \cite{Opik1957a,Moffitt1957,obrien1969,Bacci1975,Bates1978,Judd1984} and several recent \cite{Chen2011,Plotnikova2016,Liu2019a,Nikolaev2018a,Ishikawa2019,Paramekanti2020,Khaliullin2021,Mosca2021} papers but particularly mention an unduly rarely cited paper by K.D. Warren~\cite{Warren1982b}. Using the so-called angular overlap model, Warren was able to treat limiting situations of small and very large SOC for all possible occupations of $d$ electrons. It was shown that this interaction may substantially modify JT coupling constants. In a recent study a more general situation of the SOC of arbitrary strength was considered by a very different approach \cite{Streltsov2020} for $E=\{Q_2,Q_3\}$ type distortions in case of a static JT problem. 

The main results of  \cite{Streltsov2020} can be summarized as following:
Vibronic and spin-orbit coupling interactions can either enhance or suppress each other depending on a particular situation, first of all, on the number of electrons per site. For one electron at triply degenerate $t_{2g}$ states, for which the SOC is not quenched, an increase in SOC gradually suppresses the JT effect, which, however, remains nonzero even for very strong SOC. For the $d^2$ configuration, the JT effect is also suppressed by SOC. In the case of $d^4$ and $d^5$ configurations (with all electrons in $t_{2g}$ states, which is the case for the low-spin states typical of 4$d$ and 5$d$ systems) the JT effect also vanishes as the strength of SOC grows. However, in contrast with the $d^2$ case, the JT effect disappears at finite value of $\lambda_{cr}$ in an almost abrupt way, and it is strictly zero above $\lambda_{cr}$. Nevertheless, there may also be  an opposite effect: the SOC may enhance rather than suppress the JT distortions.  This is the situation for the $d^3$ configuration, for which the SOC does not impede but activates the JT effect, which for this configuration is absent for $\lambda=0$.

These results were obtained in Ref. \cite{Streltsov2020} by considering the JT coupling of $t_{2g}$ electrons with doubly degenerate $E$ (tetragonal and orthorhombic) distortions -- the so called $t \otimes E$ problem. However, the $t_{2g}$ states can be also split by trigonal distortions -- the $t \otimes T$ problem and moreover JT coupling constant for $T$ vibrations can be as large as for $E$ phonons~\cite{Iwahara2018}. 

It is both interesting scientifically and important practically to know  how the SOC would affect such trigonal distortions, which are often present in real situations. Theoretically it is even more interesting: for very strong SOC the states of $t_{2g}$ electrons are split into $j=3/2$ quartet and $j=1/2$ doublet, with  $j=3/2$ states lying lower. Such quartet is actually  formed by two Kramers doublets, i.e. the situation in this sense resembles that of the usual $e_g$ states and sometimes indeed these doublets are regarded as an effective $e_g$ orbitals, see, e.g., Ref.~\cite{Takayama2021}, but how far does this analogy go? In particular, $e_g$ levels are not split by the trigonal deformation, i.e. there is no interaction with trigonal $T$-phonons. However, the situation with $t_{2g}$ electrons in case of strong SOC might be very different from the $e \otimes E$ case, just because of a strong spin-orbit entanglement introduced by SOC. And indeed, it is known that  for $d^1$ configuration in the case of infinitely strong SOC, where we are dealing with the $j=3/2$ quartet, JT coupling to $T_2$ vibrations still survives \cite{Moffitt1957,Judd1984}.  The case of intermediate JT coupling, not considered in the previous literature, present special interest because real $4d$ and $5d$ systems usually belong to this category. This is what is done in the present paper. Another  question we concentrate on is what  the situation is with the $t \otimes T$ problem in case of strong SOC  for other electron configurations - $d^2$, $d^3$ etc. In these cases, besides purely JT electron--phonon interaction, an electron--electron interaction, especially the Hund's rule exchange, play crucial role and can strongly modify the behavior of a system, in particular the manifestations of JT effect in those.

Finally, one important comment has to be made. In Ref.~\cite{Streltsov2020} and in the present paper we mainly have in mind concentrated systems,  especially those with $4d$ and $5d$ transition metal ions, such as Kitaev magnets containing e.g. Ru$^{3+}$ (RuCl$_3$) or Ir$^{4+}$ (Li$_2$IrO$_3$), or double perovskites like Na$_2$BaOsO$_6$ or Sr$_2$CaIrO$_6$. Interplay of strong SOC and JT effect in these systems present an interesting and important problem, both theoretically and experimentally~\cite{Ishikawa2019,Streltsov2020,Kloss2021,Kim2019,Huang2022,Prodan2021,Yakui2021,Zhang2022}. To treat it one has to start from the case of single transition metal ions, as we do in the present paper. In treating the single site case some extra complications can enter the game, such as vibronic effects connected with treating not only electrons but also the lattice (nucleus) quantum-mechanically. These effects can lead to certain modifications, for example to the well-known Ham’s reduction of different nondiagonal matrix elements~\cite{Ham1965}. In concentrated system such as those we have in mind, the vibronic effects are practically always neglected, and the lattice is treated (quasi)classically, see e.g. \cite{englman1972,Keimer2020}. Keeping in mind this situation and aiming at application to concentrated system, we in our calculations  do not also take into account vibronic effects leaving the study of their possible effects for future.

\section{Model}
The model Hamiltonian used in the present paper includes three components
\begin{eqnarray}
\label{H}
\hat H = \hat H_{SOC} + \hat H_{JT} + \hat H_{U},
\end{eqnarray}
where the first, second, and third terms correspond to the SOC, the JT electron-lattice coupling, and the Hubbard on-site electron--electron interaction, respectively. The SOC is taken in a full vector form
\begin{eqnarray}
\label{SOC}
\hat H_{SOC} = -\zeta \sum_{\alpha} \hat {\bf l}_{\alpha} \cdot \hat {\bf s}_{\alpha},
\end{eqnarray}
where $\hat {\bf l}_{\alpha}$ and $\hat {\bf s}_{\alpha} $ are orbital and spin operators of the $\alpha$th electron, $\zeta$ is the SOC constant, and the minus sign appears because we deal with the $t_{2g}$ orbitals with effective orbital moment $l_{\it eff}=1$ \cite{Abragam}.  In the LS coupling scheme (for SOC weaker than the Hund's rule coupling), one can also write this part of the Hamiltonian as $H_{SOC} = -\lambda \hat {\bf L} \cdot \hat {\bf S}$, where $\hat {\bf L} =\sum_{\alpha} \hat  {\bf l}_{\alpha}$, $\hat {\bf S} =\sum_{\alpha} \hat {\bf s}_{\alpha}$ are the total orbital and spin moments of a particular configuration, and $\lambda = \zeta/2S$.

The interaction part is written in the standard rotationally invariant form \cite{Georges2013}
\begin{eqnarray}
\label{U}
 \hat H_{U} = (U-3J_H) \frac{\hat N (\hat N -1)}2 - 2J_H {\hat S}^2
 -\frac {J_H}2 {\hat L}^2 + \frac 52 J_H \hat N ,
\end{eqnarray}
where $U$ is the Hubbard repulsion (not important here since we consider a single site), $J_H$ is the Hund's rule intraatomic exchange, and $\hat N$ is the operator for the total number of electrons.

\begin{figure}[t!]
   \centering
  \includegraphics[width=0.4\textwidth]{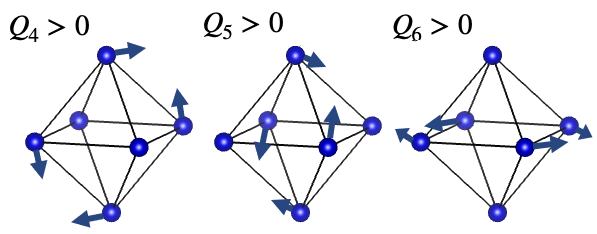}
  \caption{\label{Fig:Q4Q5Q6} Sketch demonstrating distortions of metal-ligand octahedron by trigonal $Q_4$, $Q_5$ and $Q_6$ modes.}
\end{figure}

\begin{figure}[b!]
   \centering
  \includegraphics[width=0.35\textwidth]{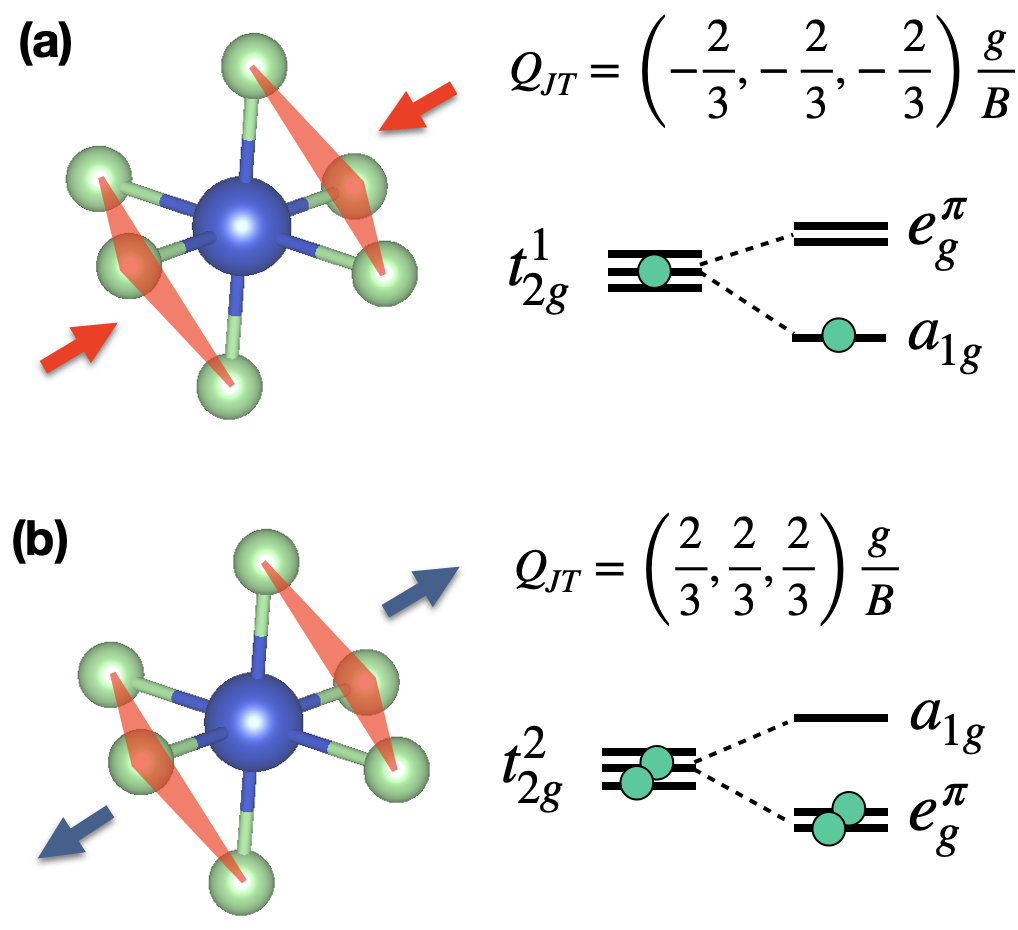}
  \caption{\label{Fig:Sketch} Energy level splitting and distortions in the $t \otimes T$ problem without the SOC in the case of $t_{2g}^1$ (a) and $t_{2g}^2$ (b) configurations.}
\end{figure}

The JT term includes the elastic energy contribution and the linear coupling of the electron subsystem with the corresponding vibrations. In Sec.~\ref{txT}, where the $t \otimes T$ problem is considered, we use the following form of the JT Hamiltonian
 \begin{eqnarray}
\hat H_{JT}^{T} &=& -g \Big( (\hat l_y \hat l_z+\hat l_z \hat l_y)Q_4 + (\hat l_x \hat l_z+ \hat l_z \hat l_x)Q_5 \nonumber \\
 &+& (\hat l_y \hat l_x+\hat l_x \hat l_y) Q_6 \Big) + \frac {B^2}2 \left( Q_4^2 +Q_5^2 +Q_6^2\right).
\label{HJT}
\end{eqnarray}
where $Q_4$, $Q_5$, and $Q_6$ are the phonon modes, illustrated in Fig.~\ref{Fig:Q4Q5Q6}, with the corresponding coefficients $g$ and $B$~\cite{Bates1978}. For simplicity, in most of numerical calculations, we assume that $g=B=1$. Positive $Q_4$, $Q_5$ and $Q_6$ in the combination $Q_4+Q_5+Q_6$ would give trigonal distortion corresponding to the elongation of octahedron in the [111] direction, see Fig.~\ref{Fig:Sketch}. A more general form the JT term used in Sec.~\ref{txd} is presented in Eq.~\eqref{Htxet}.

It has to be stressed that here we consider not a lattice of JT ions, but a single JT center. Nevertheless, we keep in mind the situation of concentrated systems.

It has to be noted that the present approach differs from conventional ones used to treat the JT effect. It is not perturbative, but it is based on numerically exact solution of the many-electron problem including all the necessary interactions (electron--lattice, Hund's rule exchange, and SOC) for an arbitrary distortion with subsequent global minimization of the total energy with respect to all possible phonon modes. In this scheme, different specific electronic states correspond to each particular nuclear configuration, i.e. all vibronic effects, such as the Ham's reduction~\cite{Ham1965,Abragam}, can be included if one would use this scheme for calculating dynamical properties of isolated JT impurities as investigated e.g. by paramagnetic resonance spectroscopy.

\section{$t\otimes T$ problem \label{txT}}

\begin{figure}[t!]
   \centering
  \includegraphics[width=0.5\textwidth]{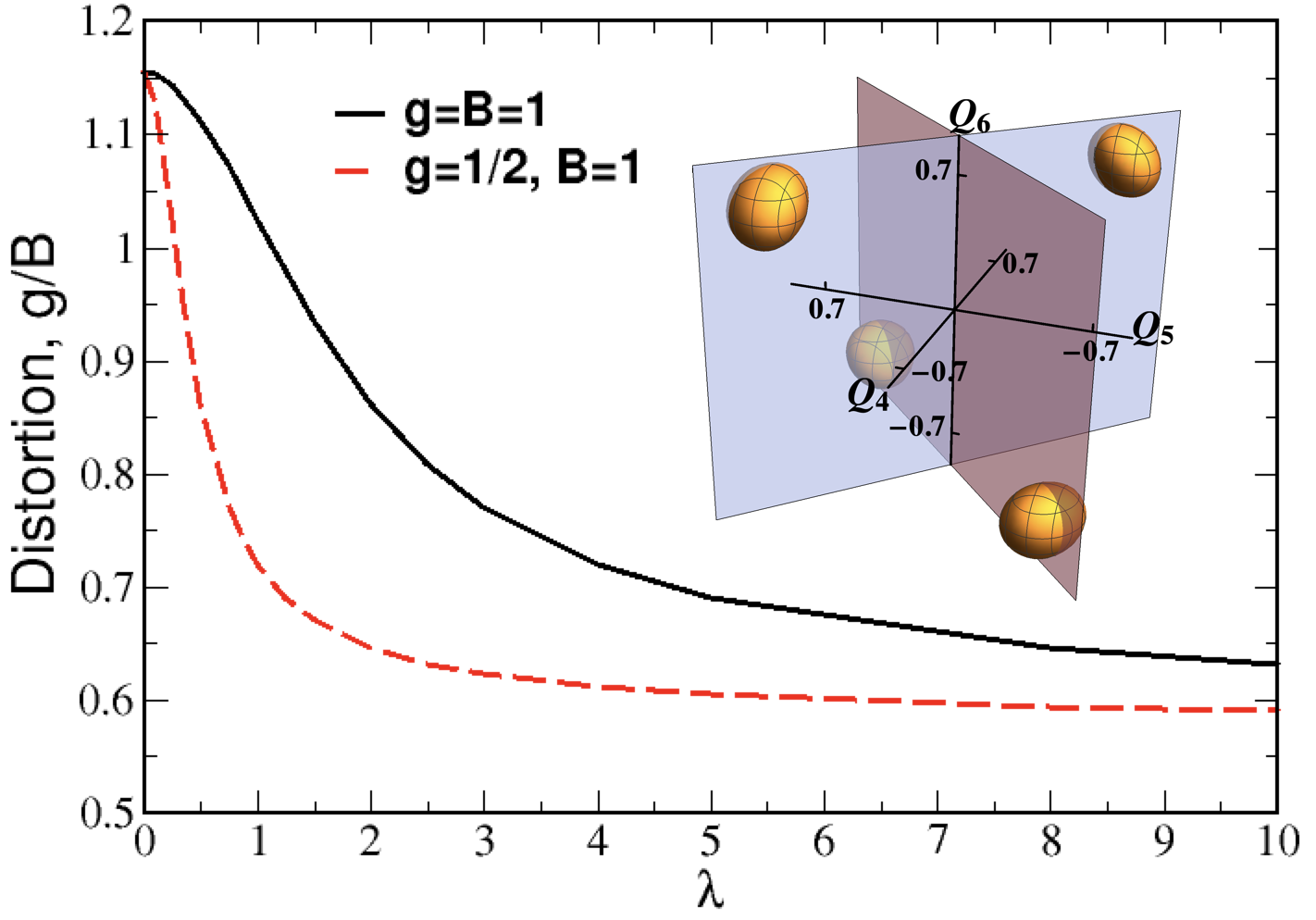}
  \caption{\label{Fig:d1-distortion} Amplitude of the JT distortion (compression) defined as $Q = \sqrt {Q^2_4 + Q^2_5 +Q^2_6}$ as a function of the SOC constant $\zeta$ in the case of $d^1$ electronic configuration for different ratios of the $g$ and $B$ parameters. The inset shows the constant energy surface for $\lambda = 0$ corresponding to $E(Q_4,Q_5,Q_6) \approx -1.3 \frac {g^2}{2B} $, which is close to the absolute energy minimum $E=-\frac 43 \frac {g^2}{2B}$.}
\end{figure}

In this section, we  not only study how the SOC affects the JT effect in the case of the $t\otimes T$ problem, but also pay attention to the role of intraatomic Hund's rule exchange. It is assumed here that the $t_{2g}-e_g$ crystal-field splitting, $10Dq$, is very large (always larger than the SOC constant $\lambda$). 

\subsection{$d^1$ configuration}

The situation in the case of $d^1$ configuration without SOC is well documented, and the JT effect results in the trigonal compression, i.e. distortion along one of four possible vectors: $[-1,-1,-1]$,  $[-1,1,1]$, $[1,-1,1]$, or $[1,1,-1]$ in the $Q_4,Q_5,Q_6$ space.   This leads to the level splitting such that the $a_{1g}$ orbital turns out to be lower than $e_g^{\pi}$ and a single electron occupies this $a_{1g}$ orbital, see Fig.~\ref{Fig:Sketch}(a). These four minima are clearly seen in the inset of Fig.~\ref{Fig:d1-distortion}, where the constant energy surface $E(Q_4,Q_5,Q_6)$ corresponding to 99\% of global energy minimum $E=-\frac 43 \frac {g^2}{2B}$ is presented. In the ($Q_4,Q_5,Q_6$) space, these minima are located along four [111] directions with the total tetrahedral symmetry. These minima would be located at other ends of these [111] axes for the opposite sign of the coupling constant $g$ in Hamiltonian \eqref{HJT}.

\begin{figure}[t]
  \includegraphics[width=0.4\textwidth]{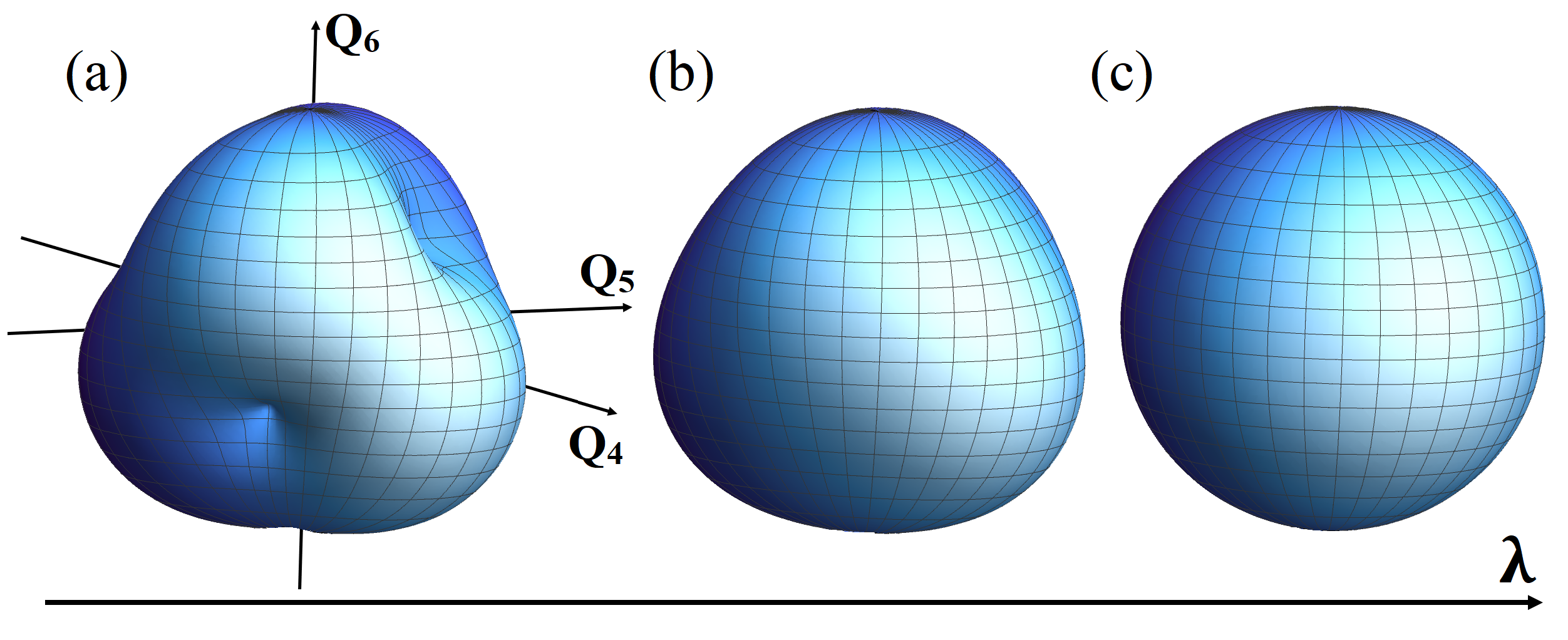}
  \caption{\label{Fig:Mexican-globe} Evolution of the energy isosurface $E(Q_4,Q_5,Q_6)$ as a function of the SOC constant $\lambda$ in the case of $d^1$ configurations. a) $\lambda=0$ b) $\lambda=5g^2/B$  c) $\lambda=50g^2/B$. To calculate the isosurface the Hamiltonian (\ref{HJT}) is rewritten in a spherical coordinates $Q_4=r \sin{\theta}\cos{\phi}$, $Q_5=r \sin{\theta}\sin{\phi}$, $Q_6=r \cos{\theta}$ and then it is minimized in $r$ for each $(\theta, \phi)$ point. }
\end{figure}

The account taken of the SOC results in the gradual suppression of the JT distortions as shown in Fig.~\ref{Fig:d1-distortion}. In agreement with previous studies the system retains trigonal minima in the presence of not very strong SOC\cite{Opik1957a,Bacci1975}. However, one might see that the SOC tends to stabilize an electron at very different orbitals (as compared to those favorable with respect to the JT effect) and this results in the suppression of the amplitude of the distortion and JT coupling constant as was noted in \cite{Warren1982b}. However, the SOC cannot lift the degeneracy completely -- we still have an electron at the doubly degenerate (without taking into account the Kramers degeneracy) $j =3/2$ subshell. Therefore, the JT effect will never be suppressed completely.

The results of these calculations also answer the question formulated in the Introduction: : To which extent does the ground-state quartet $j=3/2$ (two Kramers doublets), reached for strong SOC, resembles the $e_g$ quartet (also two Kramers doublets) for the usual $d$ electrons in cubic crystal field without SOC? We remind that for the usual $d$ electrons with one electron (as, e.g., in Mn$^{3+}$) or one hole (as in Cu$^{2+}$) at $e_g$ levels, the JT effect leading to the lifting of this degeneracy exists for tetragonal and orthorhombic distortions but not for trigonal ones.  In our case, however, for one electron at the $j=3/2$ quartet, not only tetragonal \cite{Streltsov2020} but also trigonal distortions lead to the JT effect, see Fig.~\ref{Fig:d1-distortion}. Thus we see that the $j=3/2$ quartet is in this sense not equivalent to the usual $e_g$ case. A different character of the corresponding wave functions in this case, with the strong spin-orbit entanglement, leads to different characteristics of the JT effect for strong SOC. The study of the coupling to trigonal modes (the $t \otimes T$ problem) thus allows us to reveal the role of spin-orbit entanglement for the JT effect.

\begin{figure}[t!]
  \includegraphics[width=0.45\textwidth]{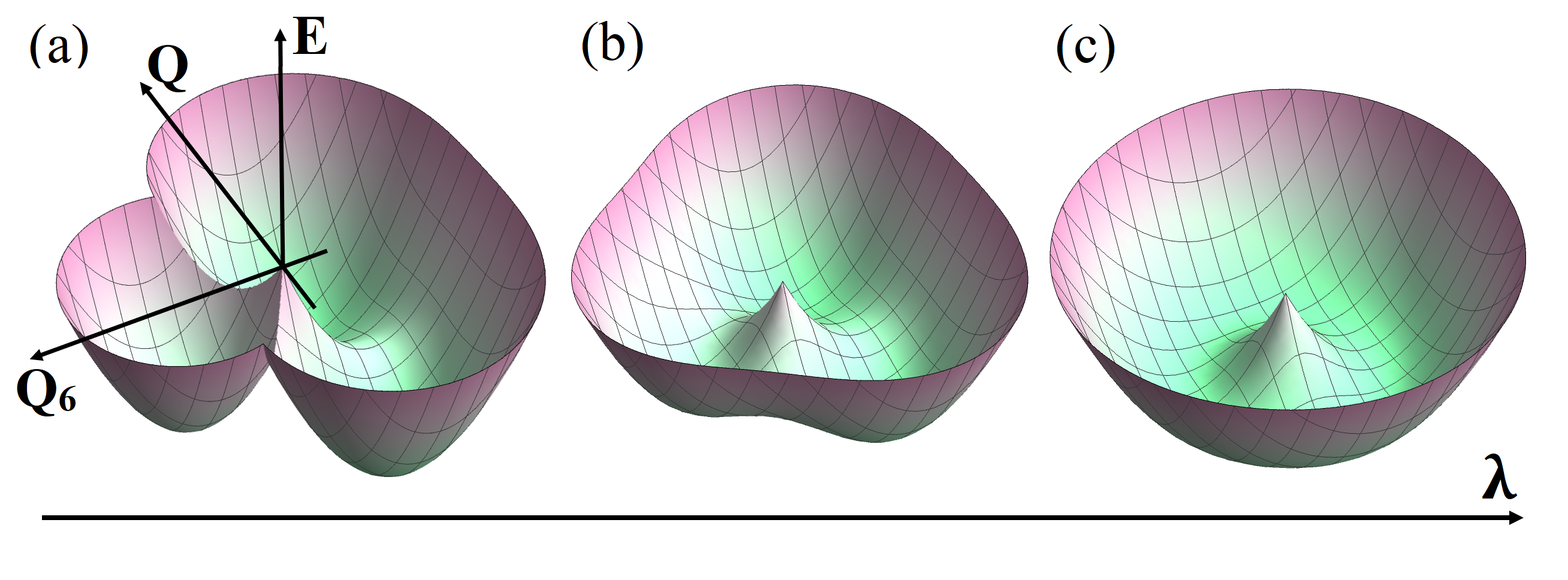}
  \caption{\label{Fig:Mexican-cut} Cuts of the ``Mexican globe'' along the $\phi=45^{\circ}$ meridian in the case of $d^1$ configurations at the SOC constant $\lambda=0$ (a), $\lambda=5g^2/B$ (b) and  $\lambda=50g^2/B$ (c). $Q = (Q_4 + Q_5)/\sqrt2$ is a normalized distortion along the $\phi=45^{\circ}$ meridian.}
\end{figure}

\begin{figure}[b!]
   \centering
  \includegraphics[width=0.45\textwidth]{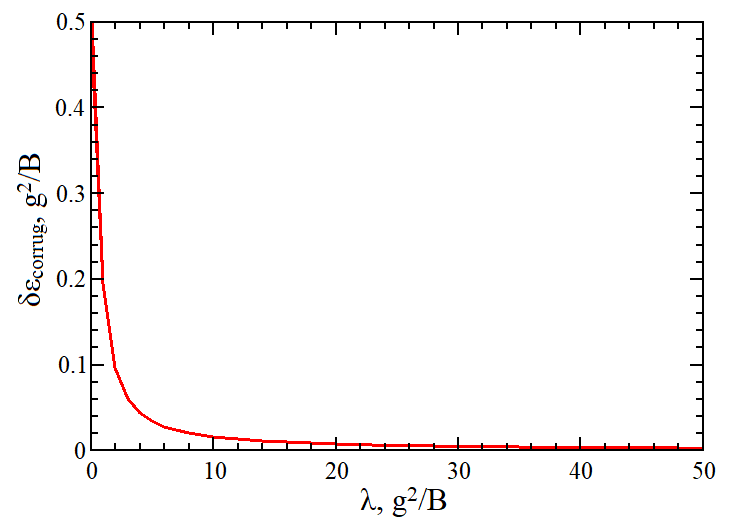}
  \caption{\label{Fig:DEcorrug} Corrugation energy $\delta \varepsilon_{\text{corrug}}$ as a function of the SOC constant $\lambda$ in the case of $d^1$ configurations. $\delta \varepsilon_{\text{corrug}}$ is the absolute difference of the energy minimum and a saddle point.}
\end{figure}

In fact, we see that for the $d^1$ configuration, the situation for trigonal distortions is actually similar to that for tetragonal ones \cite{Streltsov2020}: the strong SOC reduces JT distortions but not completely. Similarly to the $t \otimes E$ case, for the strong SOC, different trigonal distortions (elongation and compression along different [111] axes) become equivalent, so that any linear combination thereof has the same energy, which  is the situation of the ``Mexican hat''  (continuous manifold of degenerate states, which for the $e \otimes E$ problem indeed has a form of a ``Mexican hat'', see, e.g., Refs.~ \cite{Bersuker2006,khomskii2014transition,Streltsov2020-chem}).  Thus, in this case for very strong SOC, we also have the manifold of degenerate states, forming ``Mexican hat'', but in the 4D space, see Appendix~\ref{App1}. The fact that in the limit of $\lambda \to \infty$ the JT effect for the case of one electron at the $j=3/2$ quartet (in the Bethe notation, the $\Gamma_8$ quartet) leads to the continuum of degenerate states (1D manifold, the trough in the Mexican hat in the $t \otimes E$ problem, 2D manifold – the ``Mexican globe'' for the $t \otimes T$ case) is well-known in the JT literature \cite{Moffitt1957,Bates1978,Judd1984}. We demonstrated how this state is reached with increasing $\lambda$, i.e., how the energy surface evolves from that of $\lambda=0$ to the limiting solution of the Mexican hat in the $t \otimes E$ or the Mexican globe for the $t \otimes T$ case for infinite SOC.

\begin{figure}[t]
   \centering
  \includegraphics[width=0.5\textwidth]{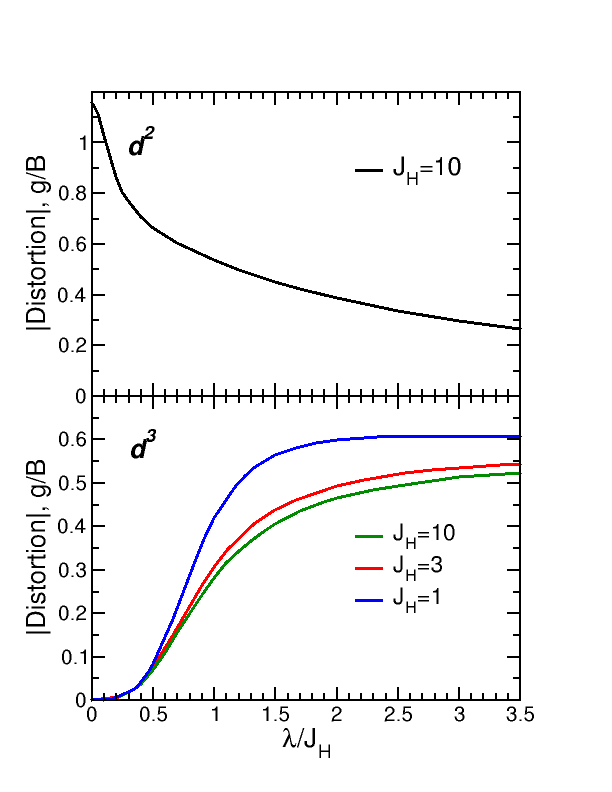}
  \caption{\label{Fig:d2-distortion} Amplitude of the JT distortion defined as $Q = \sqrt {Q^2_4 + Q^2_5 +Q^2_6}$ as a function of the SOC constant $\lambda$ in the case of $d^2$ and $d^3$  electronic configurations for $g=B=1$.}
\end{figure}

Analysis of the energy surface of the $t \otimes T$ problem is quite difficult compared to that for the  $t \otimes E$ and $e \otimes E$ problems, since $E(Q_4,Q_5,Q_6)$ is a 4D function. The energy isosurfaces $E(Q_4,Q_5,Q_6)=\text{const}$ can be plotted (e.g., see Fig.~\ref{Fig:Mexican-globe}) but they are difficult to compare with  the Mexican hats in the $t \otimes E$ and $e \otimes E$ problems. However, one can decrease dimensionality of $E(Q_4,Q_5,Q_6)$ if we make some cut of the ``Mexican globe'' by combining two phonon modes (e.g. $Q_4$ and $Q_5$) into one $Q= (Q_4 + Q_5)/\sqrt2$ (where $1/\sqrt2$ is normalization factor). It corresponds to cutting the ``globe'' by the corresponding meridian. Then we can plot  the energy surfaces of the trigonal and tetragonal modes. 

The cuts of the ``Mexican globe'' along the $\phi=45^{\circ}$ meridian at the various values of the SOC constant $\lambda$ are shown in Fig.~\ref{Fig:Mexican-cut}. Two global minima of the energy cut corresponds to the $[-1,-1,-1$] and $[1,1,-1]$ minima in the $Q_4,Q_5,Q_6$ space. Other minima of the Mexican globe (Fig.~\ref{Fig:Mexican-globe}) can be obtained if the cut is made along other directions. The last (local) minimum is in fact a saddle point in the $Q_4,Q_5,Q_6$ space; its energy is equal to the energy of the saddle point between the global minima. The energy difference between the global minima and the saddle point (with the local minima) becomes smaller with increasing SOC (see Fig.~\ref{Fig:DEcorrug}). Thus the cut of the Mexican globe turns into the well-known Mexican hat at $\lambda \gg E_{JT}$ (Fig.~\ref{Fig:Mexican-cut}c).

The cuts in Fig.~\ref{Fig:Mexican-cut} are generally similar to the Mexican hats of the $t\otimes E$ and $e\otimes E$ problems. Both such ``hats'' have conical points at $(0,0)$. The cuts have three minima in Fig.~\ref{Fig:Mexican-cut}(a) and Fig.~\ref{Fig:Mexican-cut}(b); however, whereas in the $t\otimes E$ and $e\otimes E$ problems all three minima are equal, here, in this cross-section, we get two global and one local minima. Also we obtain the continuous set of minimum points in  Fig.~\ref{Fig:Mexican-cut}(c). Except for the presence of the local minimum, the evolution of the cuts of Mexican globe in the $t \otimes T$ problem and the Mexican hat of the $t \otimes E$ problem as the function of $\lambda$ very similar. Note that in the $e \otimes E$ problem  the Mexican hat gets corrugated due to higher order JT coupling. Here, however, we get some corrugation {\it already for the linear JT coupling} but for finite SOC; so, instead of the continuous set of minimum points, only four minima exist for finite $\lambda$.

\subsection{$d^2$ configuration}

In the case of $d^2$ electronic configuration, one needs to take into account the intraatomic exchange interaction, $J_H$. Here, we assume that the energy gain due to the JT distortions is always smaller than $J_H$, but the strength of the SOC, $\lambda$, can be larger or smaller than the Hund's rule energy $J_H$ and $g^2/{2B}$.

Having two electrons in the absence of the SOC, we gain more energy by elongating the metal--ligand octahedron in [111] direction and by putting two electrons with parallel spins onto $e_{g}^{\pi}$ orbitals, as shown in Fig.~\ref{Fig:Sketch}(b). This results in the trigonal elongation along one of four possible [111] directions discussed above (in solids, the exchange interaction or electron--phonon coupling could choose one of these directions). The SOC in turn favors the occupation of very different orbitals. These are $j=3/2$ spin-orbitals, see Eq.~\eqref{3/2-states} in Appendix~\ref{App2}.

Therefore, by increasing the strength of SOC, we reduce the maximum possible energy gain (and the distortion as a result) due to the vibronic coupling, see the upper panel of Fig.~\ref{Fig:d2-distortion}.  This is exactly what is observed in our numerical calculations --  the JT distortion amplitude decreases with $\lambda$.

Moreover, formally as $\lambda \to \infty$, the JT distortions asymptotically vanish. This is in contrast with the situation with $d^1$ configuration. One can easily understand this by noting that also at very strong SOC, the intra-atomic exchange makes two electrons to occupy  $j^z_{3/2}$ and $j^z_{1/2}$, or $j^z_{-3/2}$ and $j^z_{-1/2}$, orbitals (to have maximal spin projections, see Eq.~\eqref{3/2-states}). However, the distortions induced by such occupation compensate each other exactly: Using Eqs.~\eqref{HJT} and \eqref{3/2-states}, it can be readily shown that e.g. $j^z_{3/2}$ gives $Q_{JT}^{3/2} = -\frac 13 \frac gB$, while $j^z_{1/2}$ results in $Q_{JT}^{1/2} = \frac 13 \frac gB$.

These results were obtained taking into consideration only the $t_{2g}$ manifold, assuming that the cubic crystal field leading to splitting ($10Dq$) of $t_{2g}$ and $e_g$ levels  is the largest parameter in the system. Admixture of $e_g$ states in case of finite $10Dq$  can bring about some modifications, largely of numerical character. This problem, especially important for $d^2$ configuration, will be considered separately.

\subsection{$d^3$ configuration}
In the case of three electrons and zero SOC, we fill three $t_{2g}$ levels by electrons, which have the same spin projection due to the strong intraatomic Hund's rule exchange. Such a state does not exhibit any orbital degeneracy and therefore it is inactive for the JT effect.

\begin{figure}[t]
   \centering
  \includegraphics[width=0.5\textwidth]{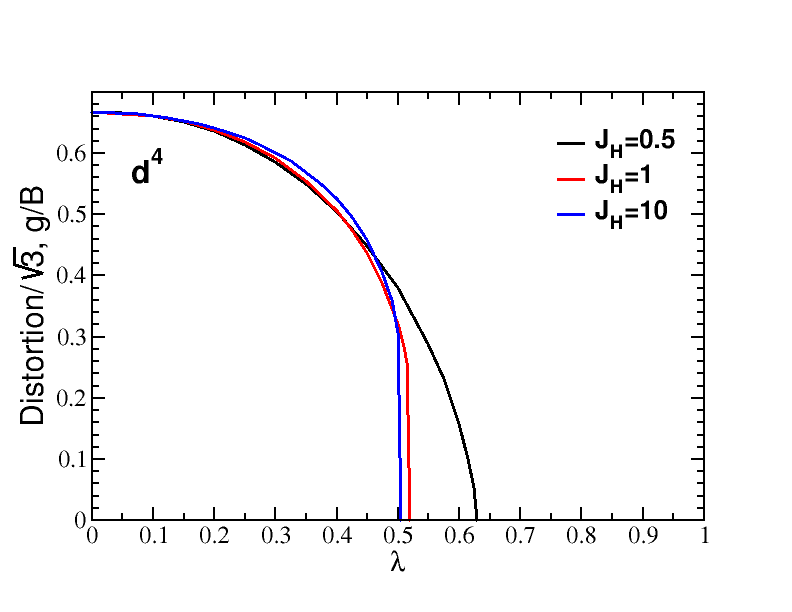}
  \caption{\label{Fig:d4-distortion} Amplitude of the JT distortion defined as $Q = \sqrt {Q^2_4 + Q^2_5 +Q^2_6}$ as a function of the SOC constant $\lambda$ in the case of $d^4$  electronic configuration for $g=B=1$.}
\end{figure}

The SOC acts against the Hund's rule exchange and redistributes electrons in such a way as to make them occupy $j=3/2$ states. This results in orbital degeneracy (three electrons at the fourfold degenerate $j=3/2$ states) and activates the JT effect  as was pointed out by Warren~\cite{Warren1982b}. The distortion amplitude as a function of SOC $\lambda$ is shown in Fig.~\ref{Fig:d2-distortion}. One might expect that, similarly to the $t \otimes E$ problem~\cite{Streltsov2020}, here one would also have the Mexican hat geometry of the adiabatic potential energy surface in the formal limit of $\lambda \to \infty$. Indeed, as in the case of the $d^1$ configuration, we have here a single ``JT active'' particle at the $j_{3/2}$ levels, but it is a hole in the case of $d^3$.

It is interesting to study the effect of intra-atomic exchange on JT distortions. First, one may see in Fig.~\ref{Fig:d2-distortion} that it is ratio $\lambda/J_H$, which plays a crucial role. A half maximum possible JT distortion is achieved at $\lambda/J_H \sim 0.7 - 1$. On the other hand, indeed the Hund's rule and SOC favor very different occupations of the spin-orbitals by electrons: eigenfunctions \eqref{eq:j32} of spin-orbit operator \eqref{SOC} are obviously not optimum  from the viewpoint of intra-atomic exchange, which favors having as many as possible electrons with the same spin projection. Therefore, by increasing $J_H$, we suppress the JT distortions induced by the SOC, see Fig.~\ref{Fig:d2-distortion}.

\subsection{$d^4$ and $d^5$ configurations}

\begin{figure}[t]
   \centering
  \includegraphics[width=0.45\textwidth]{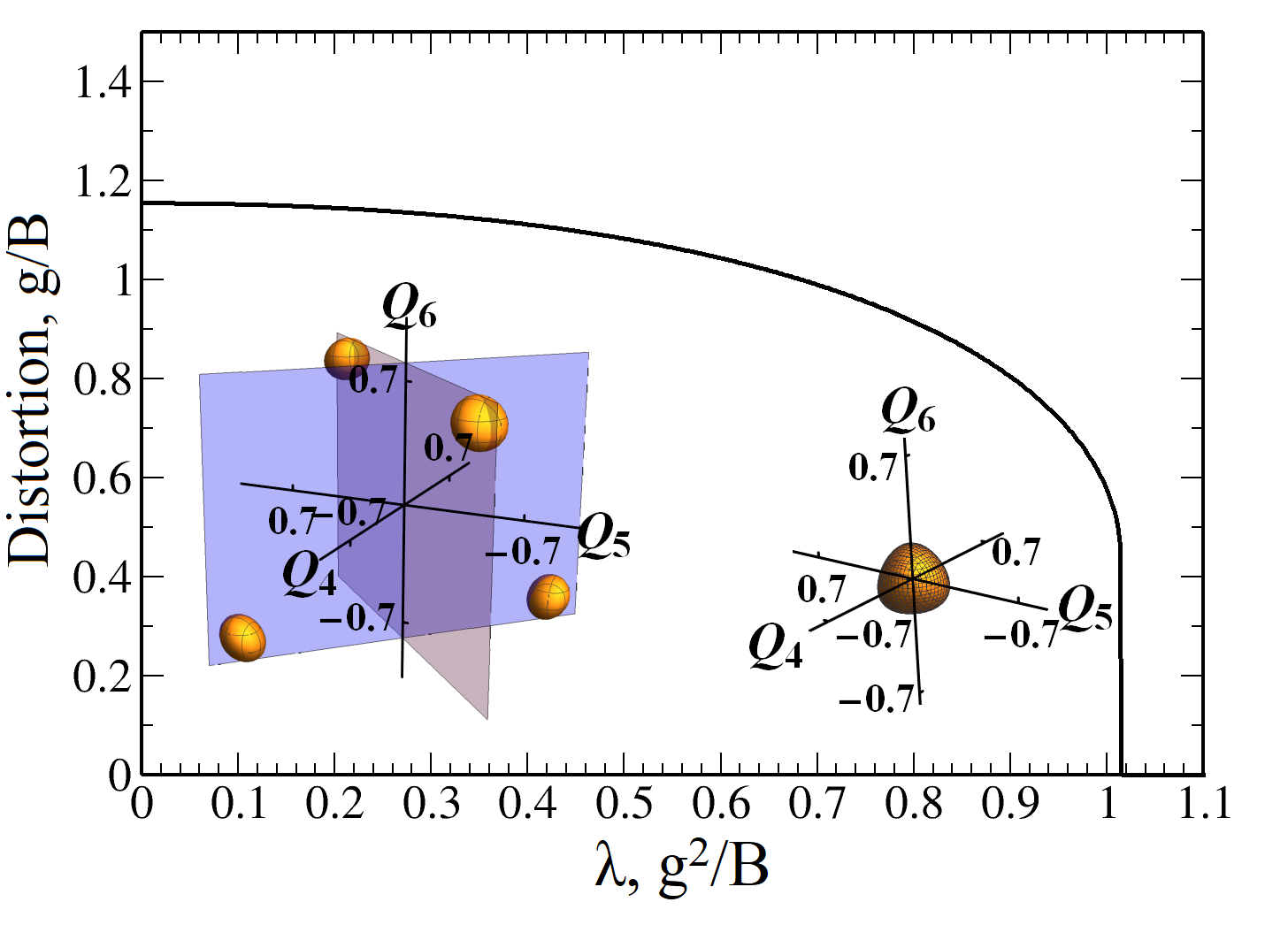}
  \caption{\label{Fig:d5-distortion} Amplitude of the JT distortion defined as $Q = \sqrt {Q^2_4 + Q^2_5 + Q^2_6}$ as a function of the SOC constant $\lambda$ in the case of $d^5$ configuration. This function is the same for different $g$ and $B$, if $\lambda$ is measured in the units of $\frac{g^2}{B}$ and $Q$ is measured in the units of $\frac{g}{B}$. Insets demonstrate the constant energy surfaces $E(Q_4, Q_5, Q_6)=E_{iso}$, which are close to the global energy minima, at $\lambda=0<\lambda_c$ (left) and $\lambda=1.2>\lambda_c$ (right).}
\end{figure}

These two configurations demonstrate very similar behavior in the case of the $ t\otimes E$ problem~\cite{Streltsov2020}. For $T$ vibrations, this result remains the same. The corresponding plots of distortion amplitude are summarized in Figs.~\ref{Fig:d4-distortion} and \ref{Fig:d5-distortion}.

The case of $d^5$ configuration without SOC can be described in terms of one hole at $t_{2g}$ levels, so it can be derived from the $d^1$ case by replacing $g$ with $-g$.  Then, the points characterizing absolute energy minima are of the opposite signs as compared to the $d^1$ case: $[1,1,1]$,  $[1,-1,-1]$, $[-1,1,-1]$, or $[-1,-1,1]$ in $Q_4 Q_5 Q_6$ space (Fig.~\ref{Fig:d5-distortion}, left inset). The $ML_6$ octahedron is elongated in one of these directions, and we put five electrons at the crystal-field levels of Fig.~\ref{Fig:Sketch}(b). This is the typical situation for such ions as Ir$^{4+}$ and Ru$^{3+}$, important, e.g., for the Kitaev materials.

As  $\lambda$ increases, the JT distortions decrease and near the critical value  $\lambda_c$ abruptly disappear (Fig.~\ref{Fig:d5-distortion}). The absence of distortions at large $\lambda$ values can be explained in the $jj$ coupling scheme, which is relevant in this case. In this scheme, a single hole occupies the upper  Kramers doublet  $j=\frac{1}{2}$, which does not have any orbital degeneracy, so the JT effect does not work in this situation.

The same picture also explains similar behavior for the $d^4$ configuration, Fig.~\ref{Fig:d5-distortion}: in the $jj$ scheme, four electrons completely fill four $j=3/2$ states leading to the nondegenerate and nonmagnetic $J=0$ state .

\section{Full $t \otimes (T + E)$ problem\label{txd}}
\begin{figure}[t]
   \centering
  \includegraphics[width=0.45\textwidth]{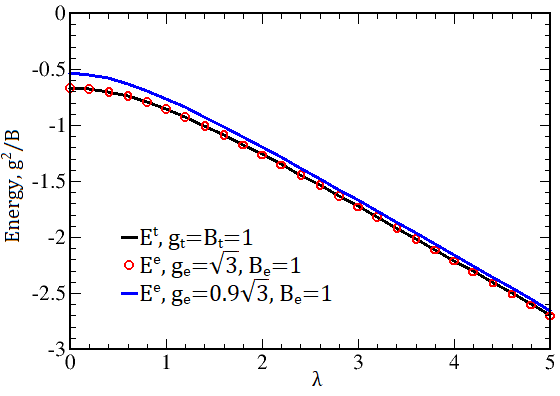}
  \caption{\label{Fig:EnergyTxet} Energy as a function of $\lambda$ for the $ t\otimes T$ (black solid line) and $ t\otimes E$ (red circles and blue solid line) problems in the case of $d^1$ configuration. The coefficient $g_e=\sqrt3$ is chosen to  have $E^e=E^t$ at $\lambda=0$ and e.g. for $g_e=0.9\sqrt3$: $|E^e|<|E^t|$.}
\end{figure}

Typically, for the description of specific materials it is enough to treat the coupling of electrons with $E$ or $T$ vibrations. Nevertheless, for completeness, below we consider the general $t \otimes (T + E)$ problem, which includes both tetragonal ($Q_2$, $Q_3$) and trigonal ($Q_4$, $Q_5$, $Q_6$) displacements. In this situation, the JT term is written in the following form 
\begin{eqnarray}
\hat{H}_{JT}^{TE}&=& \frac{B_e}{2} (Q_2^2+Q_3^2) + \frac{B_t}{2} (Q_4^2+Q_5^2+Q_6^2) \nonumber \\
&-&g_e \Big(\frac{1}{\sqrt 3}(\hat l^2_x-\hat l^2_y)Q_2+\Big(\hat l^2_z-\frac{2}{3}\Big)Q_3\Big) \label{Htxet} \\
&-&g_t\Big((\hat l_y \hat l_z + \hat l_z \hat l_y)Q_4 + (\hat l_x \hat l_z + \hat l_z \hat l_x)Q_5+  \nonumber\\
&+& (\hat l_x \hat l_y + \hat l_y \hat l_x)Q_6\Big). \nonumber
\end{eqnarray}
Here, $B_e$ and $g_e$ ($B_t$ and $g_t$) are constants corresponding to $E$ ($T$) distortions.

The solution of Eq.~\eqref{Htxet} is well known for the case of zero SOC, $\zeta=0$. There are three types of extremum points: three correspond to tetragonal minima with $Q_4=Q_5=Q_6=0$, four are trigonal minima with $Q_2=Q_3=0$, and six are orthorhombic points  \cite{obrien1969}. The difference between the energies $E^e=-2g_e^2/9B_e$ (the coupling to $E$ modes) and $E^t=-2g_t^2/3B_t$ ($T$ modes) is crucial for the $t \otimes (T + E)$ problem. If $E^e<E^t$, the tetragonal extremum points are absolute minima and the others are saddle points. Conversely, if $E^t_{JT}<E^e_{JT}$, then the trigonal points correspond to global minima, and again the others are saddle points. Orthorhombic points always remain to be saddle points.

The $E^e=E^t$ case is more complicated. All types of extrema become minimum points. Moreover, there is a continuous subset of minima. For a special case $B_e=B_t$ and $g_e=g_t/\sqrt3$ (the so-called $ t\otimes D$ problem) all minima obey the relationship $Q_2^2+Q_3^2+Q_4^2+Q_5^2+Q_6^2=Q_0^2=g_t^2/3B_t$, so they can be parameterized as
\begin{eqnarray}
Q_2&=&-\sqrt 3 Q_0 \sin^2\theta \cos2\phi, \nonumber \\
Q_3&=&-Q_0 (3 \cos^2\theta - 1), \nonumber \\
Q_4&=&-\sqrt 3 Q_0 \sin2\theta  \sin\phi, \label{Qparametrized}\\
Q_5&=&-\sqrt 3 Q_0 \sin2\theta \cos\phi, \nonumber \\
Q_6&=&-\sqrt 3 Q_0 \sin^2\theta  \sin2\phi. \nonumber
\end{eqnarray}

Let us consider how the situation changes with the account taken of the SOC. First, if we compare the results of the $ t\otimes T$ and $ t\otimes E$  problems, one can notice that all modes have similar dependence on $\lambda$.

Therefore, one might expect that the ground state energies of the $t\otimes E$ and $t\otimes T$ problems have the same dependence on $\lambda$. Direct numerical calculations of $E^e$ with $B_t=g_t=0$ and $E^t$ with $B_e=g_e=0$ (using Eq.~(\ref{Htxet})) show that this is indeed the case (see Fig.~\ref{Fig:EnergyTxet}). If $E^e$ is equal  (larger or less than)  $E^t$ at $\lambda=0$, then $E^e$ is equal (larger or less than) $E^t$ at any $\lambda$. Consequently, all conclusions derived for the $t \otimes (T + E)$ problem without SOC remain the same in the case of nonzero SOC.

Moreover, parametrization (\ref{Qparametrized})  can also be used for the case  of $B_e=B_t$, $g_e=g_t/\sqrt3$, but now all modes (including $Q_0$) become functions of $\lambda$, which are similar to the $Q$ function (Fig.~\ref{Fig:d1-distortion}). Direct calculations with the JT term described by Eq.~\eqref{Htxet} with substitution from Eq.~\eqref{Qparametrized} for some values of $\lambda$ show that the ground state energy of the system is the same for any $\theta$ and $\phi$. Thus, SOC does not destroy the continuous set of minimum points in the $ t\otimes D$ problem.

The $\lambda\to\infty$ case is considered separately. We use the same algorithm as described in Appendix~\ref{App1}, but with an additional step. Exact expression for the total energy is rather cumbersome, but one can expand this into the Laurent series at $\lambda\to\infty$ and take the leading terms ($\lambda^1$ and $\lambda^0$). After these transformations, the ground energy takes the form
\begin{equation}
 \begin{gathered}
    E(Q_2,Q_3,Q_4,Q_5,Q_6)=-\frac{\lambda}{2}
    +\frac{B_e}{2}(Q^2_2+Q^2_3)+ \frac{B_t}{2}(Q^2_4+ \\ +Q^2_5+Q^2_6) - \frac{1}{3} \sqrt{g_e^2( Q_2^2+ Q_3^2)+3 g_t^2 (Q_4^2+Q_5^2+Q_6^2)}.
 \end{gathered}
\label{Mexictxet}
\end{equation}

If we take $B_e=B_t=B$ and $g_e=g_t/\sqrt3=g$, Eq.\eqref{Mexictxet} becomes $E=B\sum_i Q_i^2 -\frac{1}{3} g \sqrt{\sum Q_i^2}-\frac{\lambda}{2}$. Now we can again use Eq.~\eqref{Qparametrized}  and obtain $E=5B Q_0^2/2 -\frac{\sqrt5}{3} g Q_0-\frac{\lambda}{2}$. The last expression has the minimum at $Q_0^2=g^2/45B^2$. This is the ``Mexican hat'' again, but in the 6D space.

Consider the case $E^e\neq E^t$. Now Eq.~\eqref{Mexictxet} depends on two sums: the sum of $E_g$ modes $Q_2^2+Q_3^2$ and the sum of $T_{2g}$ modes $Q_4^2+Q_5^2+Q_6^2$. One can transform $E_g$ modes to cylindrical coordinates and $T_{2g}$ modes to spherical coordinates
\begin{eqnarray}
Q_2&=&Q_e \sin\phi, \nonumber\\
Q_3&=&Q_e \cos\phi, \nonumber\\
Q_4&=&Q_t \sin\theta\cos\phi \label{Convert}\\
Q_5&=&Q_t \sin\theta\sin\phi, \nonumber\\
Q_6&=&Q_t \cos\theta. \nonumber
\end{eqnarray}
Then
\begin{equation}
E(Q_e,Q_t)=\frac{B_e}{2}Q_e^2+\frac{B_t}{2}Q_t^2- \frac{1}{3} \sqrt{g_e^2 Q_e^2 +3 g_t^2 Q_t^2}-\frac{\lambda}{2}.
\label{MexicQeQt}
\end{equation}
This energy equation is a 3D one, so it can be easily treated analytically. Using the first derivatives, we obtain stationary points (0, $\pm g_t/\sqrt 3 B_t$) (with $E^t_{JT}=-g_t^2/18 B_t$) and ($\pm g_e/3 B_e$, 0)  (with $E^e_{JT}=-g_e^2/6 B_e$) in $(Q_e, Q_t)$ coordinates. Then, we calculate the Hessians and find that the point (0, $\pm g_t/\sqrt 3 B_t$) is the absolute minimum if $3g_t^2 B_t>g_e^2/B_e$ (or $E^t_{JT}<E^e_{JT}$), and the point ($\pm g_e/3 B_e$, 0) is the absolute minimum if $3g_t^2 B_t<g_e^2/B_e$ (or $E^e_{JT}<E^t_{JT}$).

Hence, in the $\lambda\to\infty$ limit, we have three ``Mexican hats'': the first is 4D with only $T_{2g}$ modes ($Q_4^2+Q_5^2+Q_6^2=Q_t^2=g_t^2/3B_t$) for $E^t_{JT}<E^e_{JT}$. The second 3D hat includes only  $E_{g}$ modes $Q_2^2+Q_3^2=Q_e^2=g_e^2/9 B_e^2$ for $E^e_{JT}<E^t_{JT}$. Finally, the last ``Mexican hat'' is a 6D in $T_{2g}$ and $E_g$ modes space that was considered at the beginning of this section.

 Here we have considered the  static $t \otimes (T + E)$ problem of the linear JT coupling at arbitrary SOC strength. Previously, the dynamical properties  for the strong JT coupling \cite{obrien1969} and coexistence of tetragonal, orthorhombic and trigonal distortions for the quadratic JT interaction \cite{Bacci1975} was studied  for this problem. However, the SOC was treated only for special cases, such as $g$-value \cite{obrien1969} and $^3T$ term \cite{Bacci1975}.

\section{Conclusions}

In this paper, we analyzed an interplay between  SOC and vibronic interactions in ions with partially occupied $t_{2g}$ levels. A special emphasis was put on the $t \otimes T$ problem, i.e. on the interactions of $t_{2g}$ electrons with trigonal vibrational modes $Q_4$, $Q_5$, and $Q_6$ of a metal--ligand octahedron.

In the case of the $d^1$ configuration, an increase in the SOC leads to a gradual decay (but not vanishing) of the characteristic JT distortions. At a strong SOC, we obtain a 4D analog  of the ``Mexican hat'' adiabatic potential energy surface with the potential for concomitant quantum effects.

For the $d^2$ configuration, the SOC also suppresses the JT distortions. However, in contrast with  the $d^1$ case, these distortions can vanish due to such additional factor as the Hund's rule intraatomic exchange, $J_H$. Quite an unusual situation arises for the $d^3$ configuration, for which in the absence of SOC, owing to the strong Hund's rule exchange, three electrons with parallel spins occupy three $t_{2g}$ levels, thus removing  orbital degeneracy. The SOC redistributes such electrons favoring the occupation of  the $j = 3/2$ state. That is why, the orbital degeneracy is restored, and the JT effect begins to work, i.e. in this case the SOC does not suppress but activates JT effect.

The $d^4$ and $d^5$ cases turn out to be quite similar in their behavior. In both cases, the JT distortions abruptly vanish  at a sufficiently strong SOC since the latter favors the formation of the $j = 1/2$ doublet for the $d^4$  and a singlet $J=0$ state for $d^5$ configurations, which do not exhibit the orbital degeneracy, thus removing the JT effect.

The results, in a nutshell, are that the qualitative behavior of JT effect for trigonal distortions (the $t \otimes T$ problem) for  the strong SOC coupling is qualitatively similar to that for coupling to tetragonal distortions (the $t \otimes E$ problem) considered earlier \cite{Streltsov2020}.  This agrees with previous results, where limiting situations of a very large and small SOC were considered \cite{Warren1982b}. In particular, the JT effect for trigonal distortions can survive even for very strong SOC, when we can describe the situation by the $j=3/2$ quartet. In this sense, the situation in such a limit is not identical to the actual $e_g$ case with two Kramers doublets. The more complicated nature of the SOC-stabilized states with strong entanglement of spins and orbitals, changes the situation drastically and makes it quite nontrivial.

  An interesting conclusion is that the very strong SOC  leads to a continuous degeneracy of the ground states ("Mexican hat", in this case "Mexican globe"). In contrast with the usual situation without SOC, where this continuous degeneracy is lifted by higher-order effects, here it is destroyed already for linear JT coupling but for finite SOC. This can lead to interesting effects in particular in the dynamics  of such systems.

It is worthwhile to note that features of local distortions of the ligand octahedra in $4d$ and $5d$ transition metal compounds have become a subject of many recent studies. These are e.g. local point symmetry breaking in Ba$_2$NaOsO$_6$ seen by local methods such as NMR~\cite{Lu2017}, while diffraction does not detect any deviations from the cubic symmetry~\cite{Erickson2007} or noncubic crystal-field seen by the resonant inelastic x-ray scattering in various iridates expected to be in undistorted octachedra in the limit of large SOC ~\cite{Sala2014,Liu2012Kat,Revelli2019a}. One might also mention unexpected elongation of the octahedra in Ba$_2$SmMoO$_6$~\cite{Mclaughlin2008}, Ba$_2$NdMoO$_6$~\cite{Cussen2006}, Sr$_2$MgReO$_6$~\cite{Sarapulova2015}, Sr$_2$LiOsO$_6$~\cite{Woodward2022}, and K$_2$TaCl$_6$~\cite{Ishikawa2019}, which sometimes is accompanied by even further lowering of the symmetry and thus might involve $T$ modes.  The coupling to the trigonal vibrations should be especially relevant for systems containing corresponding transition metal ions with  face-sharing  octahedra -- such as, for example, systems of the type of  Ba$_3$TMRu$_2$O$_9$ or Ba$_3$TMIr$_2$O$_9$, with TM =Na, Ca, Y, Ce etc. Detailed study of these materials is an important, but at the same time complicated problem, since there are many other factors affecting lattice distortions in addition to the conventional JT effect such as purely steric factors defined by the Goldschmidt tolerance factor, or possible high-order multipolar orderings~\cite{Mosca2021,Pourovskii2021}.

Thus, we see that even a single-site problem involving the SOC provides a real cornucopia of interesting new physics.  Taking into consideration the interactions between JT sites in
a lattice results in an interplay among orbital, spin, and lattice degrees of freedom. It has been shown on example of the $E$ distortions that the spin-orbit and vibronic interactions also compete in this case as well and, e.g., for $d^1$ configuration suppression of the JT distortions by the SOC also occurs ~\cite{Streltsov2020}. taking into account the interactions of the JT ion with the lattice may bring even a richer physical content. Indeed, these are not simple electronic orbitals, but spin-orbitals, which are now coupled with lattice distortions and therefore one might expect other novel, e.g. magneto-elastic, effects in this case, but details depend on a particular occupation of $d$ orbitals, lattice connectivity and of course the strength of the SOC. We believe that the results of this work should create a good basis for a further study of these effects.

\section*{Acknowledgements}
S.V.S., F.T., and K.I.K. acknowledge the support of the Russian Science Foundation (project No. 20-62-46047) in the part concerning the numerical calculations. The work of D.I.Kh. is supported by the Deutsche Forschungsgemeinschaft (project No. 277146847-CRC 1238).

\appendix

\section{Goldstone modes for $d^1$ in the case of $\lambda \to \infty$ \label{App1}}
 
In this Appendix, we will show analytically that the adiabatic potential energy surface for electronic configuration $d^1$ in the limit of $\lambda \to \infty$ is similar to the ``Mexican hat'' in the space of 4D ($Q_4$, $Q_5$, $Q_6$, $E$), where $E$ is the energy.

For the sake of simplicity, we first perform the derivation for the $t \otimes E$ problem,
which was considered in detail in Ref.~\cite{Streltsov2020}. In this case, instead of $Q_4$, $Q_5$, and $Q_6$ one has only two phonon modes, $Q_2$ and $Q_3$. As the first step, we transform full Hamiltonian (3) of Ref.~\cite{Streltsov2020} including both SOC and JT terms to the basis, which is diagonal in the space of $j_{1/2}$ and $j_{3/2}$ states. In the limit of $\lambda \to \infty$, the splitting $j_{1/2}$ and $j_{3/2}$ becomes infinitely large and one can work only with $4 \times 4$ Hamiltonian for $j_{3/2}$ states. Its diagonalization gives the spectrum with the lowest in energy eigenvalue
\begin{eqnarray}
E(Q_2, Q_3) = -\frac{\lambda} 2 - \frac g3 \sqrt{Q_2^2 + Q_3^2} + \frac B2 (Q_2^2 + Q_3^2).
\end{eqnarray}
There are two types of extrema -- the first one at $(Q_2=0,Q_3=0)$ is absolutely unstable and the second one corresponding to the absolute minimum is parametrized by the equation
\begin{eqnarray}
Q_2^2 + Q_3^2 = \frac 49 \frac {g^2}{B^2}.
\end{eqnarray}
This is nothing else, but the equation describing the trough of the ``Mexican hat''. We see that the ground state of our problem is highly degenerate and it is described by the rotation in the $Q_2Q_3$ space, i.e. by the Goldstone mode.

Now, one can repeat the same calculations for the $t \otimes T$ problem. Then, we obtain
\begin{eqnarray}
E(Q_4, Q_5, Q_6) &=& -\frac{\lambda} 2 - \frac g{\sqrt 3} \sqrt{Q_4^2 + Q_5^2 + Q_6^2} \nonumber \\
 &+& \frac B2 (Q_4^2 + Q_5^2 + Q_6^2),
\end{eqnarray}
i.e. the same quadratic form characteristic for the Goldstone modes, which again gives equation for the trough of the ``Mexican hat'', but now in the 4D space
\begin{eqnarray}
Q_4^2 + Q_5^2 + Q_6^2 = \frac 43 \frac {g^2}{B^2}.
\end{eqnarray}

\section{ Wave functions \label{App2}}

If one considers a metal-ligand octahedron with the axes directed to the metal-ligand bonds, then the trigonal orbitals are
\begin{eqnarray}
\label{a1g-LCS}
|a_{1g}\rangle         &=& \frac 1 {\sqrt 3} \left( |xy\rangle +  |xz\rangle +  |yz\rangle \right), \\
|e_{g}^{\pi}\rangle &=& \pm  \frac 1 {\sqrt 3} \left( |xy\rangle +  { \rm e}^{\pm 2 \pi {\rm i}/3} |xz\rangle +   {\rm e}^{\mp 2 \pi {\rm i}/3} |yz\rangle \right).
\label{eg-LCS}
\end{eqnarray}
However, if $z$ axis is chosen along the trigonal $[1,1,1]$ direction, they can be written in a more suitable form~\cite{Khomskii2016}:
\begin{eqnarray}
\label{a1g-GCS}
|a_{1g}\rangle          &=& |3z^2-r^2\rangle , \\
|e_{g,1}^{\pi}\rangle &=& - \frac 2 {\sqrt 6}  |xy\rangle +  \frac 1 {\sqrt 3} |yz\rangle,  \nonumber \\
|e_{g,2}^{\pi}\rangle &=&  \frac 2 {\sqrt 6}  |x^2 - y^2 \rangle +  \frac 1 {\sqrt 3} |xz\rangle.
\label{eg-GCS}
\end{eqnarray}
Then, one may construct $l^z =\pm 1$ states from the $e_g^{\pi}$ orbitals
\begin{eqnarray}
|l^z_{\pm 1}\rangle  =  |e_{g,1}^{\pi} \pm {\rm i}  e_{g,2}^{\pi}\rangle,
\end{eqnarray}
while $|l^z_0\rangle  = |a_{1g}\rangle $.

Finally the $j=3/2$ wave-functions are:
\begin{eqnarray}
\label{3/2-states}
|j_{3/2}, j^z_{3/2}\rangle &=& |l^z_1, \uparrow \rangle
,\nonumber \\
|j_{3/2}, j^z_{-3/2}\rangle &=& | l^z_{-1}, \downarrow \rangle
,\nonumber \\
|j_{3/2}, j^z_{1/2}\rangle &=& \sqrt{\frac 2 3}|l^z_0, \uparrow \rangle + \frac 1 {\sqrt{3}}|l^z_{1}, \downarrow \rangle  \nonumber \\
|j_{3/2}, j^z_{-1/2}\rangle &=& \sqrt{\frac 23}|l^z_0, \downarrow \rangle + \frac 1 {\sqrt{3}}|l^z_{-1}, \uparrow \rangle
\label{eq:j32}
\end{eqnarray}

\bibliographystyle{apsrevlong_no_issn_url}
\bibliography{library2b}

\end{document}